\documentclass[12pt,a4paper]{article} 

\usepackage{epsfig}

\newcommand{\be}{\begin{equation}} 
\newcommand{\en}{\end{equation}}
\newcommand{\bea}{\begin{eqnarray}}
\newcommand{\ena}{\end{eqnarray}}

\newcommand{\hbo}{\hbox to 1 true cm {\hfill } } 
\newcommand{\tr}{\hbox{tr}}

\def\dslash{\partial\kern-.6em\slash}
\def\kslash{k\kern-.5em\slash}
\def\pslash{p\kern-.4em\slash}
\def\Dslash{D\kern-.6em\slash}
\def\Vslash{V\kern-.7em\slash}
\def\vslash{v\kern-.5em\slash}
\def\rslash{r\kern-.5em\slash}
\def\qslash{q\kern-.5em\slash}

\begin{document} 
\vglue 1truecm
  
\vbox{ UNITU-THEP-18/99
\hfill December 20, 1999
}
  
\vfil
\centerline{\large\bf The non-Abelian dual Meissner effect as 
    color--alignment } 
\centerline{\large\bf in SU(2) lattice gauge theory } 
  
\bigskip
\centerline{ Kurt Langfeld, Alexandra Sch\"afke$^*$ }
\vspace{1 true cm} 
\centerline{ Institut f\"ur Theoretische Physik, Universit\"at 
   T\"ubingen }
\centerline{D--72076 T\"ubingen, Germany}
  
\vfil
\begin{abstract}
A new gauge (m-gauge) condition is proposed by means of a generalization 
of the Maximal Abelian gauge (MAG). The new gauge admits a space time 
dependent embedding of the residual U(1) into the SU(2) gauge group. This 
embedding is characterized by a color vector $\vec{m}(x)$. It turns out 
that this vector only depends of gauge invariant parts of the link 
configurations. Our numerical results show color ferromagnetic 
correlations of the $\vec{m}(x)$ field in space-time. The correlation 
length scales towards the continuum limit. For comparison with the MAG, 
we introduce a class of gauges which smoothly interpolates between 
the MAG and the m-gauge. For a wide range of the gauge parameter, the 
vacuum decomposes into regions of aligned vectors $\vec{m}$. 
The ''neutral particle problem'' of MAG is addressed in the context 
of the new gauge class.

\end{abstract}

\vfil
\hrule width 5truecm
\vskip .2truecm
\begin{quote} 
$^*$ supported by Graduiertenkolleg {\it Struktur und Wechselwirkung 
von Hadronen und Kernen}. 

PACS: 11.15.Ha, 12.38.Aw

{\it Dual Meissner effect, confinement, Abelian projection, gauge 
fixing, lattice gauge theory } 
\end{quote}
\eject

\section{ Introduction } 
\label{sec:1} 

Revealing the mechanism which operates in Yang-Mills theories for 
confining quarks to color singlet particles is one of the most important 
tasks of modern quantum field theory. A knowledge of this mechanism 
will help to understand nuclear forces from first principles and 
will, e.g., have a strong impetus on the present understanding of 
hadron physics. 

\vskip 0.3cm 
In the recent twenty years of active research, many proposals have been 
launched to explain quark confinement (see~[1-7] for an incomplete list). 
Nowadays large scale computer simulations assign top priority 
to the so-called color superconductor mechanism~\cite{tho76}. 
This mechanism applies in the Abelian gauges~\cite{tho76}, 
which allow for a residual U(1) gauge degree of freedom. {\it Projection} 
is performed to reduce the full SU(2) to compact U(1) gauge theory. After 
this projection, monopoles which carry quantized color-magnetic charges 
with respect to the residual U(1) group naturally appear as degrees of 
freedom~\cite{kro87}. The color-superconductor mechanism operates as follows: 
a condensation of these monopoles implies a (dual) Meissner effect. 
Color-electric flux is squeezed into flux tubes implying that the potential 
between two static color sources is linearly rising at large distances. 
Modern computer facilities provide the testing grounds: evidence has been 
accumulated that the dual superconductor picture captures parts of the roots 
of quark confinement~\cite{pisa,sch99}. 

\vskip 0.3cm 
Unfortunately, the color-superconductor picture suffers from the 
following drawback: color states which are neutral with 
respect to the residual U(1) gauge are insensitive to the condensate 
of the color magnetic monopoles and therefore do not acquire a confining 
potential~\cite{gia99}. In this case, one would expect additional 
''light'' states in the particle spectrum besides hadrons and glueballs. 
This contradiction to the experiment requires a refinement of the dual 
superconductor picture. For this purpose, it was argued~\cite{gia99} that 
{\it all} color-magnetic monopoles which are 
defined by different Abelian projection schemes condense while only 
those monopoles which correspond to the gauge fixing at hand are 
manifest. The idea of the condensation of ''hidden'' monopole 
degrees of freedom might conceptually solve the ''neutral particle problem'', 
but conceals the non-Abelian nature of the superconductor mechanism. 

\vskip 0.3cm 
In this paper, we throw a new glance onto the non-Abelian Meissner 
effect by refraining from a residual U(1) gauge group which is 
uniquely embedded into the SU(2) gauge group all over space time. 
Instead of, space-time is decomposed into regions in each of which 
the embedding of the residual U(1) group into the SU(2) gauge group 
is chosen to yield the minimal error by a subsequent projection. 
Particles which are ''neutral'' with respect to the U(1) 
subgroup in one particular region carry charge 
in another region. If the average over all configurations is performed during 
the Monte-Carlo sampling, all particles are confined on distances 
which exceed the intrinsic length scales of the regions. 

\vskip 0.3cm 
For putting this idea on solid grounds, we propose a new type of 
gauge (referred to as m-gauge) which appears as a generalization of the 
Maximal Abelian gauge (MAG). In the m-gauge, the orientation of the 
U(1) subgroup in SU(2) is specified by a unit-color vector $\vec{m}$ 
which depends on space-time. In a first step, we show by means of 
numerical calculations that the m-projected theory still bears quark 
confinement at full strength. Secondly, for putting the m-gauge into a 
proper context, we investigate, by virtue of a gauge fixing parameter, a 
gauge fixing which smoothly interpolates between the MAG and the m-gauge.
For a wide range of the gauge fixing parameters, the vacuum decomposes 
into regions of (approximately) aligned color vectors~$\vec{m}$. 

\vskip 0.3cm 
The paper is organized as follows: in the next section, we will introduce 
the novel type of gauge fixing and the corresponding projection. 
The numerical errors induced by projection are studied for the case 
of MAG and the case of m-gauge, respectively. In section \ref{sec:3}, we 
reveal the color ferromagnetic correlations of the vector $\vec{m}$ 
present in the m-gauge. The interpolating gauges are introduced 
and the vacuum structure in these gauges is discussed. 
Conclusions are left to the final section.

\section{ M-gauge } 
\label{sec:2} 

\subsection{ The new gauge fixing and projection } 
\label{sec:2.1} 

A particular configuration of the SU(2) lattice gauge theory is 
represented by a set of link matrices $U_\mu (x) \, \in \, SU(2)$, which 
are transformed by a gauge transformation $\Omega (x) \, \in \, SU(2) $ 
into 
\be 
U_\mu (x) \rightarrow U^\Omega _\mu (x) \; = \; \Omega (x) \, 
U_\mu (x) \, \Omega ^\dagger (x+\mu ) \; . 
\label{eq:1} 
\en 
Our new proposal for the gauge fixing condition is given by 
\bea 
S_{\mathrm{fix}} &=& \frac{1}{2} 
\sum _{\mu , \, \{x\} } \; \tr \left\{ U^\Omega _\mu (x) \, 
m(x) \,  \left( U^\Omega \right) ^\dagger _\mu (x) \, m(x) \right\} 
\; \rightarrow \; \hbox{maximum} \; , 
\label{eq:2} \\ 
m(x) &=& m^a(x) \, \tau ^a \; , \; \; \; \vec{m}^T (x) \vec{m} (x) = 1 \; , 
\nonumber 
\ena 
where $\tau ^a$ are the SU(2) Pauli matrices. 
For a given configuration $U_\mu (x)$ we allow for a variation of the 
gauge matrices $\Omega (x) $ and of the auxiliary unit vector $m^a(x)$ 
for maximizing the functional $S_{\mathrm{fix}}$. Note that a reflection of 
the vector, i.e., $\vec{m} (x) \rightarrow - \vec{m} (x)$, does not change 
the gauge fixing functional. Identifying the points $\vec{m}$ and 
$-\vec{m}$ of the sphere $S_2$ defines a projective space $RP_2$ which carries 
the gauge fixing information. Furthermore, $S_{\mathrm{fix}}$ is invariant 
under a multiplication of the gauge matrix $\Omega $, under consideration, 
with a center element of the SU(2) gauge group, i.e., $\Omega 
\rightarrow (-1) \, \Omega $. This implies that the theory after gauge 
fixing possesses a residual $Z_2$ gauge invariance, at least. 
It turns out that a further invariance is unlikely to exist 
for generic link configurations $U_\mu (x)$ (see discussion in 
subsection~\ref{sec:2.2}). 

\vskip 0.3cm 
The concept of {\it projection} is to reduce the number of degrees 
of freedom while preserving the confining properties of SU(2) 
gauge theory. It might be easier in the reduced theory to reveal the 
mechanism of confinement than resorting to the full SU(2) gauge theory. 
In the present case, we define the projected links $\hat{U} _\mu (x) $ by 
\be 
\hat{U} _\mu (x) \; := \; N \, \left[ U^\Omega _\mu (x) \; + \; 
m(x) \, U^\Omega _\mu (x) \, m(x)  \, \right] \; , 
\label{eq:3} 
\en 
where the normalization $N$ is obtained by demanding $\hat{U} _\mu (x) \, 
\hat{U} ^\dagger _\mu (x) \, = \, 1 $. 

\vskip 0.3cm 
It is convenient for an illustration of the gauge fixing (\ref{eq:2}) and 
the projection (\ref{eq:3}) to decompose the link variable as 
\be 
U^\Omega _\mu (x) \; := \; a_\mu ^{(0)}(x) \; + \; i \, \vec{a} _\mu (x) \, 
\vec{\tau} \; , \hbo \left( a_\mu ^{(0)}(x) \right)^2 \, + \, 
\vec{a}_\mu ^{\;2}(x) \; = \; 1 \; \, \; \; \forall \mu, x
\label{eq:4} 
\en 
In this case, the gauge fixing condition $S_{\mathrm{fix}} $ (\ref{eq:2}) 
becomes 
\be 
S_{\mathrm{fix}} \; = \; \sum _{\mu , \, \{x\} } \; \left\{ 
\left( a_\mu ^{(0)}(x) \right)^2 \, - \, \vec{a}_\mu ^{\;2}(x) 
\, + \, 2 \, \left( \vec{m} \vec{a}_\mu \right)^2 \right\} 
\; \rightarrow \; \hbox{maximum} \; . 
\label{eq:5} 
\en 
Representing the vector $\vec{a}_\mu (x) $ by components parallel to 
$\vec{m}(x)$ 
and perpendicular to $\vec{m}(x)$, i.e., $\vec{a}_\mu (x) = 
(a_\mu ^{\parallel}, a_\mu ^{1\perp }, a_\mu ^{2\perp } )^T $, the 
condition (\ref{eq:5}) is equivalent to 
\be 
S_{\mathrm{fix}} \; = \; \sum _{\mu , \, \{x\} } \; \left\{ 1 \; - \; 
2 \left[ \left( a_\mu ^{1\perp } (x) \right)^2 \, + \, 
\left( a_\mu ^{2\perp } (x) \right)^2 \right] \right\} \; \rightarrow \; 
\hbox{maximum} \; . 
\label{eq:6} 
\en 
This equation tells us that the gauge fixing (\ref{eq:2}) minimizes 
the link components $\vec{a}_\mu (x)$ perpendicular to the vector 
$\vec{m}(x)$. 

\vskip 0.3cm 
Inserting (\ref{eq:4}) in (\ref{eq:3}), one finds for the projected link 
variables 
\bea 
\hat{U} _\mu (x) &=& \; 2 N \, \left[ a_\mu ^{(0)}(x) \; + \; i \, 
\vec{m}(x) \vec{a}_\mu (x) \; \vec{m}(x) \vec{\tau } \, \right] 
\label{eq:7} \\ 
&=& 2N \, \left[ a_\mu ^{(0)}(x) \; + \; i \, a_\mu ^{\parallel} (x) \; 
\vec{m}(x) \, \vec{\tau } \, \right] \; . 
\label{eq:8} 
\ena 
Projecting link configurations is a two-step process: firstly, one 
exploits the gauge degrees of freedom to minimize the link components 
$a_\mu ^{1\perp } (x)$, $a_\mu ^{2\perp } (x)$, perpendicular to 
$\vec{m}(x)$, and, secondly, these components $a_\mu ^{1\perp } (x)$, 
$a_\mu ^{2\perp } (x)$ are dropped for obtaining the 
projected link variable $\hat{U} _\mu (x)$. For quantifying the error of 
this projection, we introduce 
\be 
\omega \; := \; \frac{1}{N_{\mathrm{link}}} \, 
\left\langle S^{\mathrm{max}}_{\mathrm{fix}}[U] \right\rangle _U \; , 
\label{eq:9} 
\en 
where $N_{\mathrm{link}}$ is the number of lattice links and 
$S_{\mathrm{fix}}^{\mathrm{max}}[U]$ is 
the maximum value of the gauge fixing functional $S_{\mathrm{fix}}$ 
(\ref{eq:2}) 
for a given link configuration $U_\mu (x)$. The brackets in (\ref{eq:9}) 
denote the Monte-Carlo average over all link configurations. 
An inspection of the equations (\ref{eq:6}) and (\ref{eq:8}) tells us 
that projection yields the exact result if $\omega $ possesses the largest 
value possible, i.e., $\omega = 1$. The error increases if $\omega $ 
decreases.

\vskip 0.3cm 
Note that the space-time dependence of $m(x)$ relative to the link 
variables $U^\Omega _\mu (x)$ in the functional $S_{\mathrm{fix}} 
$ (\ref{eq:2}) is 
obtained by the demand for minimizing the error induced by projection.
This demand dictates that $m(x)$ cannot be identified with a 
so-called Higgs field which figures in general Abelian gauges~\cite{jan99}. 
For an illustration of this fact, let $m(x)$, $\Omega (x)$ be the 
configurations which maximize $S_{\mathrm{fix}}[U]$ (\ref{eq:2}) for a given 
link configuration $U_\mu(x)$, and let assume that the fields 
$m(x)$, $\Omega (x)$ are uniquely defined (this is the generic case; 
see subsection~\ref{sec:2.2}). We introduce $U^{V}_\mu (x)$ as the link 
variables which are obtained from $U_\mu(x)$ by the gauge transformation 
$V(x)$, and repeat the gauge fixing procedure with the $U^{V}_\mu (x)$ 
as basis. If $m^V(x)$, $\Omega ^V(x)$ denote the configurations which 
correspond to the maximum of $S_{\mathrm{fix}}[U^V]$, one finds 
$\Omega ^V(x) = \Omega (x) \, V(x)$ and $m^V(x)=m(x)$. The later relation 
tells us that $m(x)$ only depends on the gauge invariant parts of the 
link variables $U_\mu (x)$, and therefore encodes physical information. 
By contrast, an auxiliary Higgs fields transforms homogeneously 
under the gauge transformation $V(x)$. 

\vskip 0.3cm 
Let us compare our new gauge, defined by (\ref{eq:2}), with the 
Maximal Abelian gauge (MAG). The latter gauge can be obtained from 
the gauge condition (\ref{eq:2}) if one does not allow a variation 
of $m(x)$ with space time. 
For constant vectors $\vec{m}$, one might choose $\vec{m} $ 
to point in three direction in color space without 
a loss of generality. For an SU(2) gauge theory in four dimensions, one 
associates four link variables with each space time point, and therefore 
counts $12$ degrees of freedom at each lattice site ($9$ physical and 
$3$ gauge degrees of freedom). The MAG projection effectively reduces 
the SU(2) gauge theory to an U(1) one. After projection, the number of 
degrees of freedom per site is therefore $4$. In the new gauge, presented 
here, naive counting yields four Abelian links and the unit vector 
$\vec{m}(x)$, i.e., $6$ degrees of freedom at each site. 

\vskip 0.3cm 
We finally comment on the Gribov problem to round out this subsection. 
Note that the following remarks also apply to the class of 
general Abelian gauges, and that, in particular, the practical problem 
in implementing the gauge is not a specialty of the new gauge proposed 
in the present letter. Although the gauge fixing starting from the 
condition (\ref{eq:2}) is conceptually free of Gribov ambiguities if 
one seeks out the {\it global } maximum of the functional $S_{\mathrm{fix}}$ 
(\ref{eq:2}), one recovers the Gribov problem in practice when the 
algorithm fails to detect the global maximum. In the context of 
variational gauges, several strategies have been proposed for 
evading this problem. One possibility is introducing a Laplacian version 
of the gauge fixing condition for adapting the problem to the numerical 
capabilities. Results are available in the literature for the case of 
Landau gauge~\cite{vin92}, for the case of MAG~\cite{sij97} and for 
the case of the center gauge~\cite{for99}. Another possibility is 
to introduce quantum gauge fixing~\cite{la99} for putting the gauge fixing 
which is implemented by the algorithm in the proper context. 
Here, we will not perform a detailed study of the ''practical'' Gribov 
problem in the context of the new gauge (\ref{eq:2}). In the present first 
investigation, we will only check whether the numerical results 
(see next section) are stable against random gauge transformations 
on the link variables before invoking the gauge fixing algorithm.

\subsection{ Quality of projection } 
\label{sec:2.2}

Our numerical simulations were performed using the Wilson action and 
a lattice with $12^4$ space-time points. For $\beta $-values 
in the scaling window, i.e., $\beta \in [2.1,2.5]$, 200 heat--bath 
steps~\cite{creu80} were performed for initialization. When gauge 
fixing is requested by the application of interest, we used a standard 
iterative procedure with over-relaxation for finding the maximum value of 
the gauge fixing functional $S_{\mathrm{fix}}$ (\ref{eq:2}). 
Once we have obtained configurations $m(x)$, $\Omega (x)$ which maximize 
the functional $S_{\mathrm{fix}}$, we distort these configurations 
and re-handle the gauge fixing. Repeating this procedure (for a given 
link configuration $U_\mu (x)$) several times, we find unique 
fields $m(x)$, $\Omega (x)$ at the maximum of $S_{\mathrm{fix}}$. 
This provides numerical evidence that the (local) maximum of 
$S_{\mathrm{fix}}$ is stable against small gauge transformations, and 
that flat directions in the configuration space of $m(x)$, $\Omega (x)$ 
do not exist for generic link configurations. A thorough study of the 
Faddeev-Popov determinant in the case of m-gauge is requested for 
a rigorous proof of this fact. This is left to future work. 

\vskip 0.3cm 
In a first investigation, we calculated the Creutz ratios $\chi _{kk}$ 
with help of the expectation values of quadratic Wilson loops of 
length $r=ka$, where $a(\beta )$ is the lattice spacing and $k$ is an 
integer. It is convenient 
for the extrapolation to the continuum limit to introduce the 
scale $\Lambda $ via 
\be 
\Lambda ^2 \; = \; 0.12 \; \frac{1}{a^2(\beta )} \; \exp \left\{ - 
\frac{ 6 \pi ^2 }{11} \, \left( \beta - 2.3 \right) \; \right\} \; , 
\label{eq:10} 
\en 
which is a renormalization group invariant quantity when one-loop 
scaling applies in the asymptotic $\beta $-region. The normalization 
of $\Lambda ^2 $ is chosen for reproducing the full SU(2) string tension, 
i.e. $ \Lambda ^2 \rightarrow \sigma $ for $\beta \gg 2.1 $.

\begin{figure}[t]
\centerline{ 
\epsfxsize=9cm
\epsffile{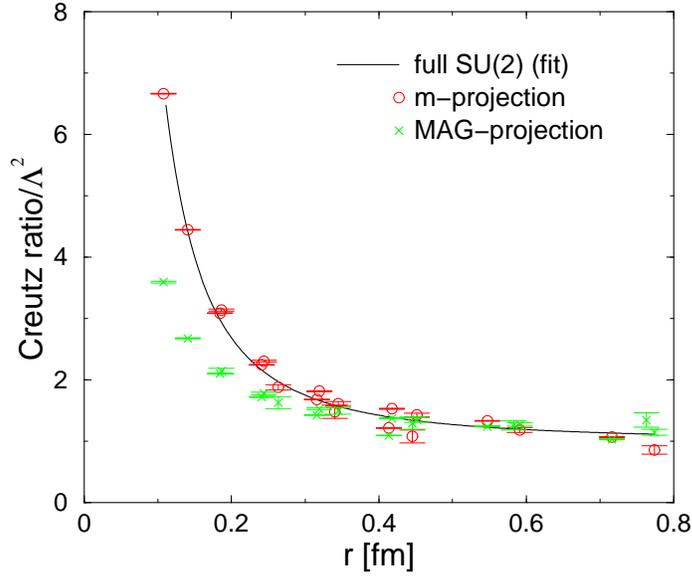}
}
\caption{ Creutz ratios: fit to data of the full theory  (solid line), 
    m-projection (circles) and MAG projection (crosses). } 
\label{fig:1} 
\end{figure}
\vskip 0.3cm 
Figure \ref{fig:1} shows our numerical data for $\chi _{kk} $ in units 
of $\Lambda ^2$ as function of $r \, \Lambda $. It turns out that 
the data of the simulation employing the full SU(2) Wilson action 
is best fitted by (solid line) 
\be 
\chi _{kk} \Lambda ^2 \; = \; \gamma _1+ \; \gamma _2 / r^2 \; . 
\label{eq:11} 
\en 
This ansatz for $\chi _{kk} $ is expected by relating Creutz ratios 
to the derivative of the static quark potential. The second term of 
the latter equation refers to the Coulomb interaction while the first 
term is present due to a non-vanishing string tension. 
Figure \ref{eq:1} also shows the data points for $\chi _{kk} $ 
calculated from links $\hat{U}_\mu (x)$ which are obtained from 
m-projection (see (\ref{eq:7})). These data are contrasted with the 
result for $\chi _{kk} $ calculated with MAG projected links. In any case, 
the asymptotic, i.e. $r \Lambda \gg 1$, value for the string tension is 
reproduced within the numerical accuracy. A striking feature of figure 
\ref{fig:1} is that the data from m-projected links agree with 
the full result for the sizes $r$ explored in figure \ref{fig:1}. 

\vskip 0.3cm 
For a quantitative study of the error induced by projection, we calculated 
$\omega $ (\ref{eq:9}) for the case of the m-projection and 
the MAG projection, respectively (see table \ref{tab:1}).
\begin{table} 
\caption{ The quality of projection, i.e., $\omega $ (\protect{\ref{eq:9}}), 
   as function of $\beta $. } 
\label{tab:1}
\vskip 0.3cm
\begin{center} 
\begin{tabular}{lccccc}
$\beta $  &  $2.1$  &  $2.2$  &  $2.3$  &  $2.4$  &  $2.5$  \\[0.5ex]
$\omega $ [m-proj.] & $0.932(6)$  & $0.934(4)$  & $0.936(3)$  & 
$0.938(3)$  & $0.940(0)$ \\ 
$\omega $ [MAG proj.] & $0.67(0)$  & $0.68(4)$  & $0.70(2)$  & 
$0.72(1)$  & $0.73(8)$  
\end{tabular} 
\end{center}
\end{table} 
One observes that $\omega $ is much bigger for m-projection rather than 
for the case of MAG projection. In particular, the components of 
the (gauge fixed) links $U^\Omega $ which are dropped by projection, i.e. 
$U^\Omega \rightarrow \hat{U}$, 
are roughly of $5\% $ in size while the generic error due to projection 
in the case of MAG is of order $30 \%$. One therefore expects that 
not only the string tension but also other physical observables which are 
calculated with m-projected links are well described. 
On one hand this feature is highly desired for constructing effective 
theories which cover a wide span of low energy properties of 
Yang-Mills theory. On the other hand, the introduction of additional 
degrees of freedom (compared with the case of MAG) obscures those ones 
which are responsible for confinement.

\section{ Color alignment in m-gauge } 
\label{sec:3}

\subsection{ M-vector correlations  } 
\label{sec:3.1} 

\begin{figure}[t]
\centerline{ 
\epsfxsize=6cm
\epsffile{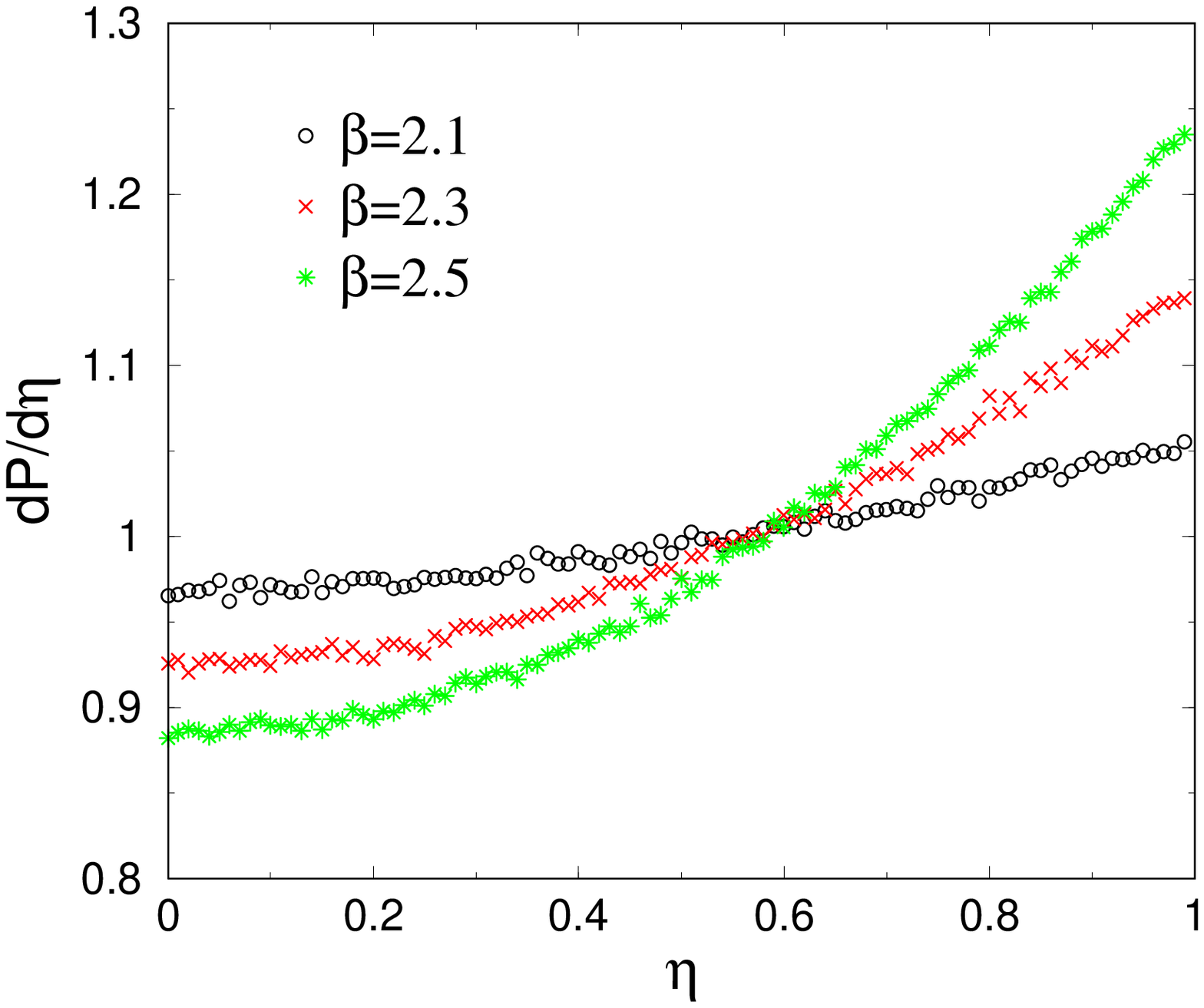}
\hspace{.5cm}
\epsfxsize=6cm
\epsffile{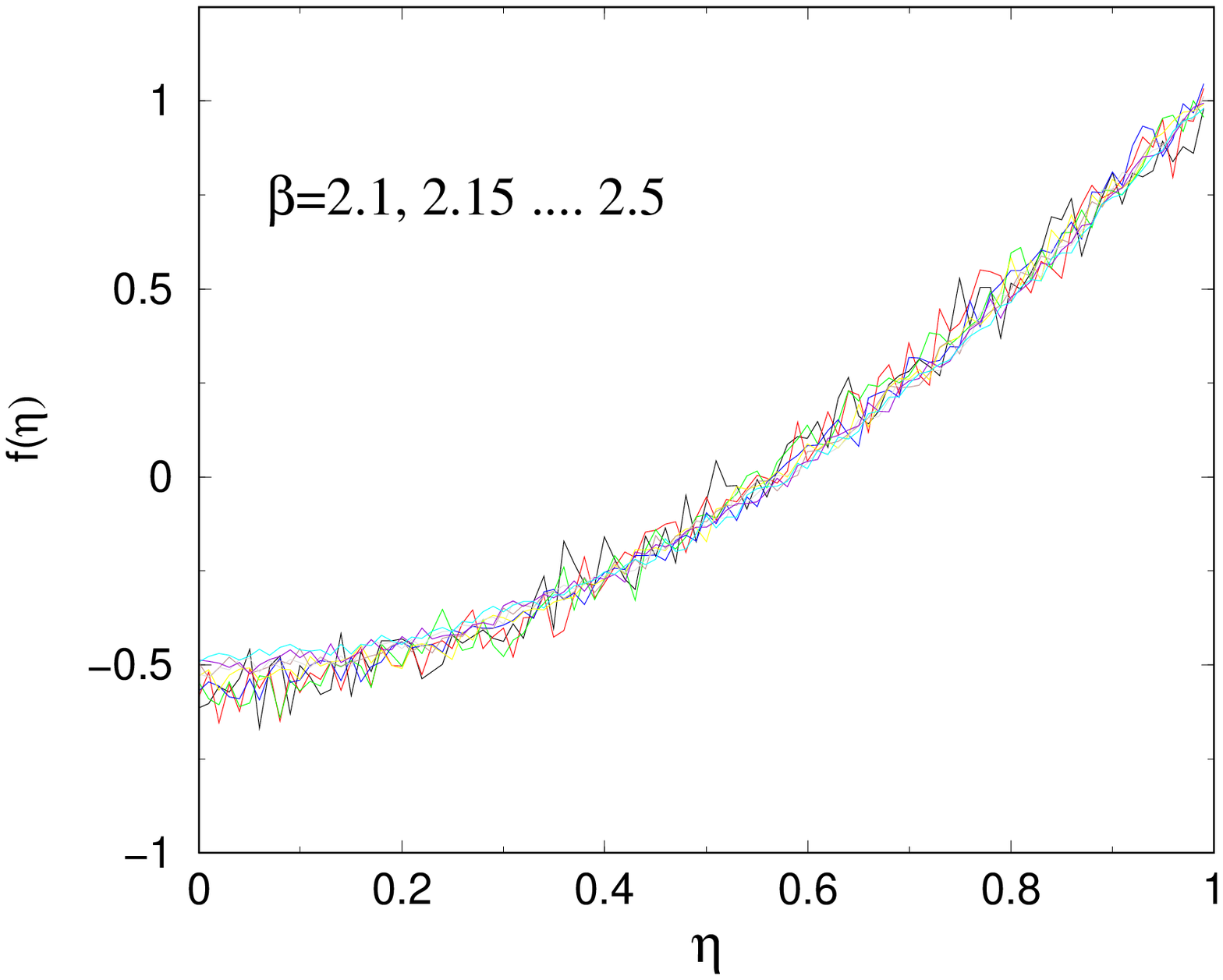}
}
\caption{ Raw data of the probability distribution of the scalar 
   product $\eta $ between two neighboring color vectors $\vec{m}$ 
   (left) and function $f(\eta )$ (right). } 
\label{fig:2} 
\end{figure}
Let us assume that we are investigating a physical observable which 
possesses a correlation length $\xi $ by comparing the full with 
the m-projected theory. If the color vector $\vec{m}(x)$ is uniquely 
oriented in a space-time domain of size $l\gg \xi $, one would recover 
the standard MAG scenario (provided that $l$ is bigger than the scale 
set by the critical temperature, i.e., $l>0.7 \, $fm, for detaining the 
Casimir effect). In particular, the dual superconductor 
picture would be expected operating if the string tension is the 
quantity of interest. For investigating the existence of such domains, 
and, in case, for relating the infra-red physics in the 
m-gauge to the well studied physics in MAG, a thorough study of the 
space-time correlations of the vectors $\vec{m}(x)$ is highly desired. 

\vskip 0.3cm
The space-time dependence of the gauge transformation $\Omega $ 
which maximizes the functional $S_{\mathrm{fix}}$ (\ref{eq:2}) induces a 
correlation of the unit vectors $\vec{m}(x)$ in space-time. 
For revealing these correlations, we numerically calculated the 
probability distribution of finding a particular scalar product $\eta $
of two vectors $\vec{m}$ located at neighboring sites of distance 
$a(\beta )$. The raw data of this distribution are shown in the left panel 
of figure \ref{fig:2}. A random distribution of vectors would correspond 
to $dP / d\eta =1 $. We clearly observe a maximum of the probability 
distribution at $\eta =1 $. We find a {\it color ferromagnetic } 
correlation between the color vectors $\vec{m}(x)$. The value at the 
maximum position increases for increasing $\beta $, i.e., for a decreasing 
distance $a(\beta )$ between the neighboring vectors. This indicates 
that $\vec{m}(x)$ which constitutes a lattice vector model so far 
will become a smooth field in the continuum limit. 

\begin{figure}[t]
\centerline{ 
\epsfxsize=9cm
\epsffile{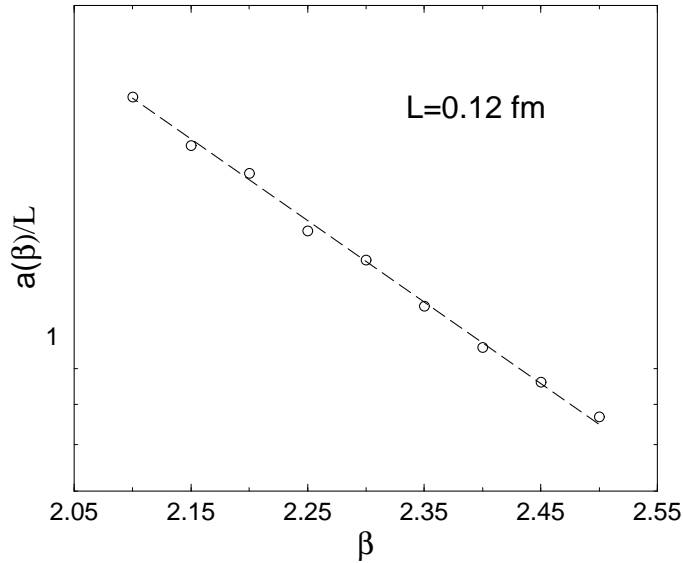}
}
\caption{ The scaling of the color ferromagnetic correlation length $L$. } 
\label{fig:3} 
\end{figure}
\vskip 0.3cm
For an interpretation of these results in the scaling limit 
$a(\beta ) \rightarrow 0$, it is useful to parameterize the 
probability distribution as follows: 
\be 
\frac{dP}{d \eta } \; = \; 1 \; + \; c \, \exp \left\{ - \frac{ 
a(\beta ) }{ L } \right\} \; f(\eta ) \; , 
\label{eq:12} 
\en 
where the function $f(\eta )$ satisfies without a loss of generality 
the constraints 
\be 
\int _0^1 f(\eta ) \; d \eta \; = \; 0 \; , \hbo f(1)=1 \; . 
\label{eq:13} 
\en 
The crucial finding is that the constants $L$ and $c$ as well as the 
function $f(\eta )$ are universal, i.e., independent of the renormalization 
point specified by $\beta $ (see figure \ref{eq:2} right panel and figure 
\ref{fig:3}). 
Comparing the numerical data for $a(\beta )/L$ with the asymptotic one-loop 
$\beta $-dependence (dashed line in figure \ref{fig:3}) 
\be 
a(\beta ) \; \approx \; 0.16 \, \hbox{fm} \, 
\exp \left\{ - \frac{3 \pi ^2}{11} \beta \right\} 
\label{eq:14} 
\en 
(where a string tension $\sigma = (440\:\mathrm{MeV})^2$ was used as reference 
scale), the extrapolation to the continuum limit yields 
\be 
L \; = \; 0.1(2) \, \hbox{fm} \; , \hbo 
c \; = \; 0.50(6) \; . 
\label{eq:15} 
\en 
Note that for observing color ferromagnetic domains a probability 
distribution $dP/d\eta (\beta \rightarrow \infty )$ 
is required which diverges at $\eta =1$. 
However, our numerical data obtained in the scaling window 
$\beta \in [2.1,2.5]$ agree with a finite value for $dP/d\eta $ at 
$\eta =1$. Further numerical investigations (e.g.~of the volume dependence 
of the distribution) are necessary for a definite conclusion on this issue.

\vskip 0.3cm 
In conclusion of this section, we find color ferromagnetic interactions 
between neighboring color vectors $\vec{m}$ which increase for 
decreasing distance $a(\beta )$ thus indicating that the vector field 
$\vec{m}(x)$ is smooth in the continuum limit. In the scaling limit, 
we find that these correlations extend over a range of roughly 
$0.12 \,$fm. The color ferromagnetic interaction between the vectors 
is (most likely) not strong enough to induce the formation of color 
ferromagnetic domains in space-time.

\subsection{ Interpolating gauges } 
\label{sec:3.2} 

\begin{figure}[t]
\centerline{ 
\epsfxsize=6cm
\epsffile{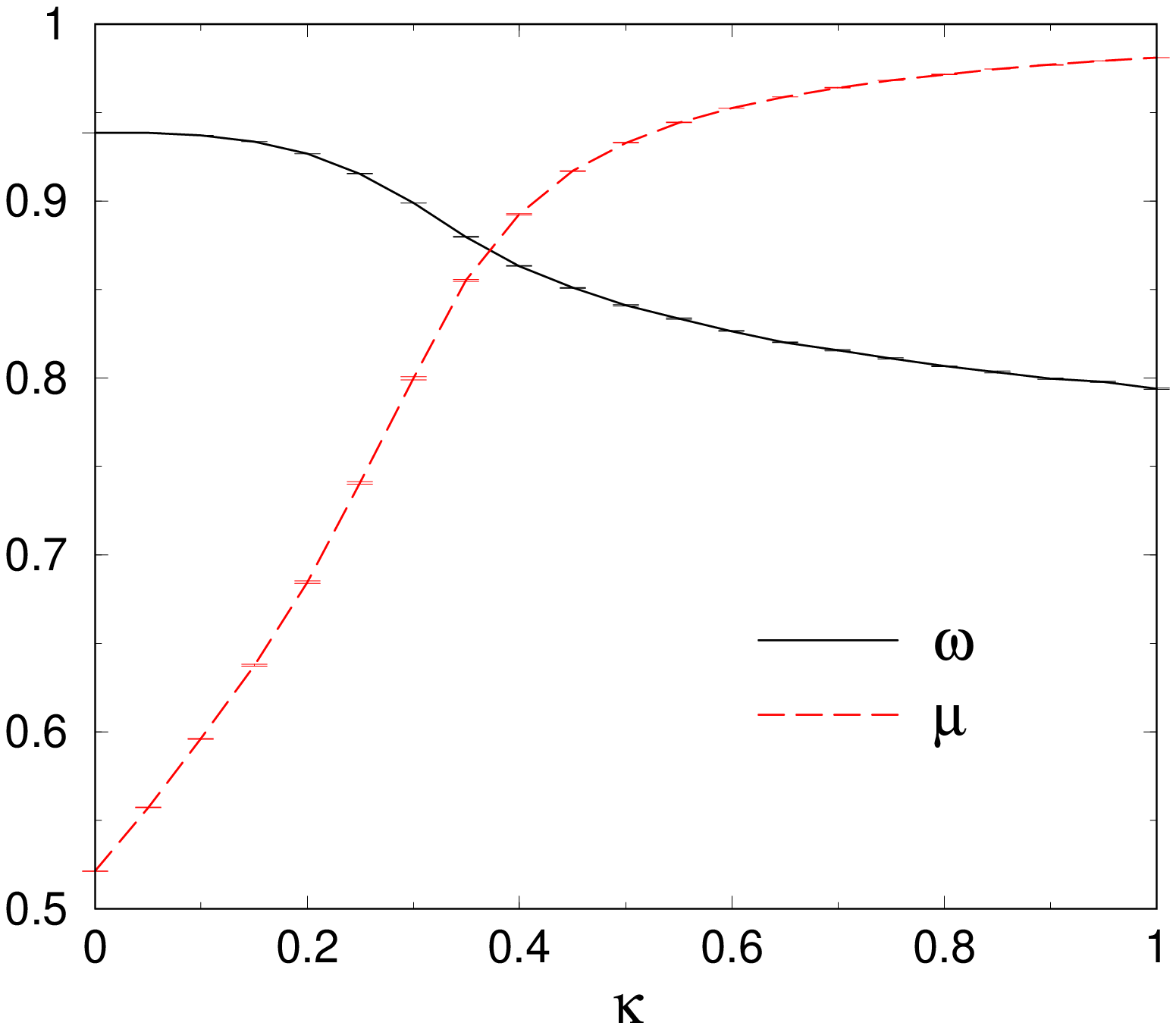}
\hspace{.5cm}
\epsfxsize=6cm
\epsffile{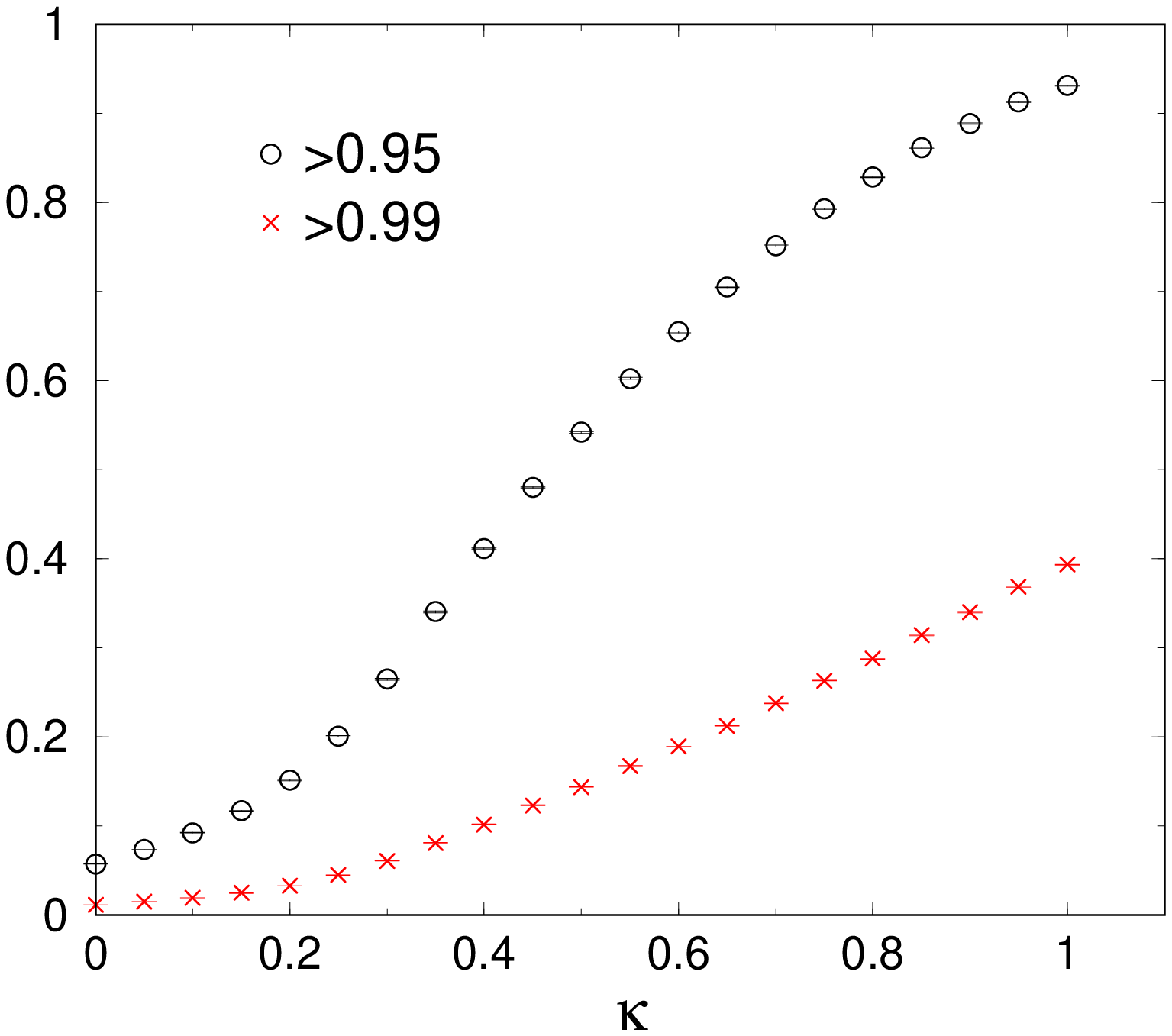}
}
\caption{ Quality of projection $\omega $ and the color ferromagnetic 
   interaction strength $\mu $ (left panel); fraction of neighboring vectors 
   with a    scalar product larger than $0.95$ and $0.99$, respectively 
   (right panel).}
\label{fig:4} 
\end{figure}
\begin{figure}[t]
\centerline{ 
\epsfig{file=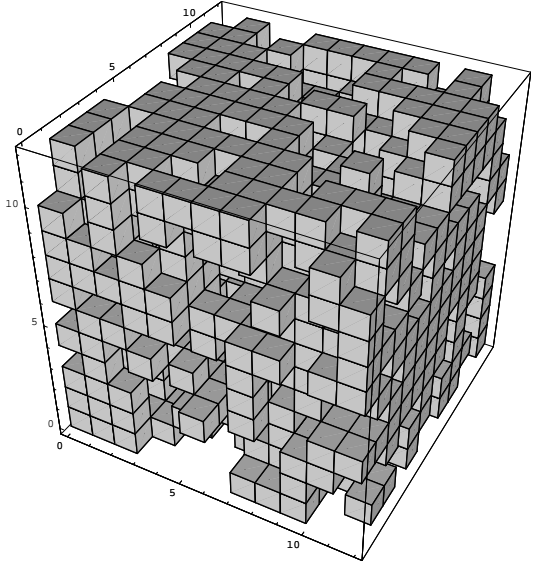,width=7cm,clip=}
}
\caption{ Spatial hypercube of one Monte-Carlo sample: region of aligned 
   color vectors $\vec{m}(x)$ for $\kappa =0.6$. } 
\label{fig:5} 
\end{figure}
Subsection \ref{sec:2.2} has demonstrated that the m-gauge is well adapted 
for projection. Unfortunately, the distribution of the auxiliary 
color vectors $\vec{m}(x)$ does not support an arrangement of these 
vectors in domains of constant orientation therefore impeding 
an interpretation of the m-gauge as local realization of MAG. 
For taking full advantage of the elaborated studies of physics 
in MAG~\cite{pisa,sch99}, we generalize the m-gauge condition (\ref{eq:2}) 
for allowing a smooth interpolation between the MAG and the m-gauge 
by virtue of a gauge fixing parameter $\kappa $. The generalized 
gauge fixing action\,\footnote{We thank Torsten Tok for helpful 
discussions on useful extensions of $S_{\mathrm{fix}}$ (\ref{eq:2}).} is 
\bea 
S_{\mathrm{fix}} &=& \frac{1}{2} 
\sum _{\mu , \, \{x\} } \; \tr \left\{ U^\Omega _\mu (x) \, 
m(x) \,  \left( U^\Omega \right) ^\dagger _\mu (x) \, m(x) \right\} 
\label{eq:16} \\ 
&+& \kappa \, \sum _{ \mu , \, \{x\} } \; \bigg[ \vec{m}(x) \, \vec{m} 
(x+\mu ) \biggr] ^2 
\; \rightarrow \; \hbox{maximum} \; , 
\label{eq:17} 
\ena 
Note that the additional term (\ref{eq:17}) also respects the 
reflection symmetry $\vec{m}(x) \rightarrow - \vec{m}(x)$. 
For $\kappa =0 $, one recovers the m-gauge (\ref{eq:2}). 
For $\kappa \gg 1$, on the other hand, there is a large penalty in action 
$S_{\mathrm{fix}}$ for non-uniformly oriented color vectors $\vec{m}$. 
One therefore 
obtains the MAG for sufficiently large $\kappa $. For quantifying the 
color ferromagnetic interaction strength, we introduce 
\be 
\mu \; := \; \frac{1}{N_{\mathrm{link}}} \, \sum _{\mu ,\, \{x\} } \; 
\biggl\langle \, \vec{m}(x) \vec{m}(x+\mu ) \, \biggr\rangle \; . 
\label{eq:18} 
\en 
One finds $\mu =1/2$ for a random distribution of $\vec{m} \in RP_2$, 
and retrieves the MAG for $\mu =1 $. 

\vskip 0.3cm 
Figure \ref{fig:4} shows our numerical results for the ''quality of 
projection'', i.e. $\omega $ (\ref{eq:9}), and 
$\mu $ as function of $\kappa $ for a $12^4$ lattice and for $\beta =2.4$. 
As expected, the strength parameter $\mu $ gradually increases with rising 
$\kappa $ while $\omega $ monotonically decreases. The minimal error 
by projection is obtained in m-gauge ($\kappa =0 $). 

\vskip 0.3cm 
The strength parameter $\mu $ at large values of $\kappa $ indicate 
that regions of uniformly oriented color vector $\vec{m}$ form. 
Figure \ref{fig:4} also shows the fraction of vector pairs  
$\vec{m}(x)$, $\vec{m}(x+\mu)$ which possess a scalar product larger 
than $0.95$ ($0.99$). The data are obtained on $12^4$ lattice 
and for $\beta =2.4$. For illustrating the regions of aligned color vectors 
at large values of $\kappa $, figure \ref{fig:5} presents the spatial 
orientation of the color vectors $\vec{m}$ for one Monte-Carlo sample at a 
given time slice. The sample was obtained for $\kappa =0.6 $. 
A reference vector was chosen at the center of the 
spatial hypercube. If the scalar product of a vector $\vec{m}$ located at 
the position $x$ with the reference vector exceeds $0.95$, an elementary 
cube which is spanned by the four points $x$, $x+\mu $, $\mu =1 \ldots 3$ 
is marked. One observes that a particular region of (approximately) 
aligned vectors $\vec{m}$ is multi-connected and extends all over the 
lattice universe. This property of the regions of alignment does not match 
with its analog in solid state physics, i.e. the Weiss domains of 
ferromagnetism.

\section{ Conclusions } 
\label{sec:4} 

In the Maximal Abelian gauge (MAG), a uniquely oriented color vector 
$\vec{m}$ 
defines the embedding of the residual U(1) into the SU(2) gauge group. 
Evidence has been accumulated~\cite{pisa,sch99} that in this case 
an (Abelian) dual Meissner effect confines particles which carry 
color-electric charge with respect to the U(1) subgroup. Since 
colored states which are, however, neutral from the viewpoint of 
the U(1) gauge group escape the confining forces provided by the dual 
superconductor mechanism, a refinement, i.e., a non-Abelian version, of the 
dual Meissner effect is highly desired. The concept of ''hidden monopoles'' 
is one possibility~\cite{gia99}. 

\vskip 0.3cm 
By generalizing the MAG gauge condition, we have here proposed 
another possibility for a non-Abelian version of the dual Meissner effect. 
The new gauge (m-gauge) admits a space-time dependent embedding, 
characterized by the color vector $\vec{m}(x)$, of the residual U(1) 
into SU(2) gauge group. The space-time dependence of $\vec{m}(x)$ is 
self-consistently chosen to achieve the minimal error induced by projection. 
It turns out that the color vector $\vec{m}(x)$ does not change 
under ''small'' gauge transformations of the link variable. Thus, 
the field $\vec{m}(x)$ carries gauge invariant information encoded 
in the link variables. Our numerical results show color ferromagnetic 
correlations of these vectors $\vec{m}$ which extends over a range of 
$\approx 0.1(2) \, $fm. The strength of these correlations seems to be 
too small for causing the formation of color ferromagnetic domains. 

\vskip 0.3cm 
For relating the m-gauge to the MAG, we have introduced a class of gauges 
which smoothly interpolates between the MAG and the m-gauge by 
virtue of a gauge fixing parameter $\kappa $. For a wide span of 
$\kappa $, the vacuum decomposes into multi-connected regions which 
are characterized by uniquely oriented vectors, and which extend all 
over the lattice universe. The internal structure of these regions 
define an intrinsic lenght scale $l_0$. Each region bears the potential 
of an Abelian Meissner effect which operates with respect to the residual 
U(1) subgroup of SU(2). Colored states which do not feel a confining force 
in one particular region generically carry charge in another sector of 
space time. We speculate that, on performing the Monte-Carlo sampling, 
all colored states are confined on length scales bigger than the intrinsic 
size $l_0$ of the regions of color alignment. Note that the size $l_0$ is 
controlled by the gauge parameter $\kappa $. The request that 
the average size is a physical quantity defines the ''running'' 
of the gauge parameter, i.e., the function $\kappa (\beta )$. 
The actual size of the regions of color alignment in physical units then 
defines the renormalized value $\kappa _R$ and must be provided by a 
renormalization condition. 

\vskip 0.3cm 
In subsumption, for a class of gauge conditions, 
specified by a gauge parameter $\kappa >0 $, the vacuum consists of 
regions of aligned color vectors $\vec{m}(x)$. The m-gauge appears as the 
limiting case $\kappa =0$. In this case, the error induced by projection 
is minimal at the expense of additional degrees of freedom as compared with 
the MAG. This fact renders the identification of the degrees of freedom 
relevant for quark confinement more difficult than in the MAG, but 
makes the m-gauge a convenient starting point for formulating an 
effective theory covering a wide span of low energy properties of 
SU(2) Yang-Mills theory.

\vskip 1cm

{\bf Acknowledgments: }
We greatly acknowledge helpful discussions with M.~Engelhardt, M.~Quandt,  
H.~Reinhardt and T.~Tok. We are indebted to H.~Reinhardt for support.


\begin {thebibliography}{sch90}
\bibitem{bak68}{ M.~Baker, J.~S.~Ball and F.~Zachariasen,
   Phys. Rev. {\bf D 51} (1995) 1968. } 
\bibitem{nie79}{ H.~B.~Nielsen and P.~Olesen, Nucl. Phys. {\bf B160}
   (1979) 380; 
   J.~Ambj{\o}rn and P.~Olesen, Nucl. Phys. {\bf B170}
   [FS1] (1980) 60; 
   J.~Ambj{\o}rn and P.~Olesen, Nucl. Phys. {\bf B170}
   [FS1] (1980) 265; 
   P.~Olesen, Nucl. Phys. {\bf B200} [FS4] (1982) 381. } 
\bibitem{adl81}{ S.~L.~Adler, Phys. Rev. {\bf D 23} (1981) 2905; 
   S.~L.~Adler and T.~Piran, Phys. Lett. {\bf B113} (1982) 405, 
   Phys. Lett. {\bf B117} (1982) 91; \\ 
   W.~Dittrich and H.~Gies, Phys. Rev. {\bf D54} (1996) 7619. } 
\bibitem{dos87}{ H.~G.~Dosch, Phys. Lett. {\bf B190} (1987) 177; 
   H.~G.~Dosch and Yu.~A.~Simonov, Phys. Lett. {\bf B205} (1988) 339. } 
\bibitem{fai88}{ G.~Fai, R.J.~Perry and L.~Wilets,
   Phys. Lett. {\bf B208} (1988) 1. } 
\bibitem{pol91}{ J.~Polonyi, {\it The Confinement and localization of quarks}, 
   In *Dobogokoe 1991, Proceedings, Effective field theories of the
   standard model* 337-357. } 
\bibitem{tho76}{ G.~'t~Hooft, {\it High energy physics }, 
   Bologna {\bf 1976}; S.~Mandelstam, Phys. Rep. {\bf C23 } (1976) 245; 
   G.~'t~Hooft, Nucl. Phys. {\bf B190} (1981) 455. } 
\bibitem{kro87}{ A.~S.~Kronfeld, G.~Schierholz, U.-J.~Wiese, 
   Nucl. Phys. {\bf B293} (1987) 461. } 
\bibitem{pisa}{ A.~Di Giacomo, B.~Lucini, L.~Montesi and G.~Paffuti,
   {\it Colour confinement and dual superconductivity of the vacuum}, 
   I and II, hep-lat/9906024, hep-lat/9906025. } 
\bibitem{sch99}{ K.~Schilling, G.S.~Bali and C.~Schlichter,
   Nucl. Phys. Proc. Suppl. {\bf 73} (1999) 638. } 
\bibitem{gia99}{ A.~Di Giacomo, Plenary talk given at 14th International 
   Conference on Ultrarelativistic Nucleus-Nucleus Collisions (QM 99), 
   Torino, Italy, May 1999, hep-lat/9907010; \\ 
   A.~Di Giacomo, B.~Lucini, L.~Montesi and G.~Paffuti, hep-lat/9906024. } 
\bibitem{jan99}{ {\it see e.g.}, O.~Jahn, hep-th/9909004. } 
\bibitem{vin92}{ J.~C.~Vink and U.~Wiese, Phys. Lett. {\bf B289} (1992) 
   122. \\ 
   J.~C.~Vink, Phys. Rev. {\bf D51} (1995) 1292. } 
\bibitem{sij97}{ A.~J.~van der Sijs, Nucl. Phys. Proc. Suppl. {\bf 53} 
   (1997) 535, Nucl. Phys. Proc. Suppl. {\bf 73} (1999) 548. } 
\bibitem{for99}{ C.~Alexandrou, M.~D'Elia and P.~de Forcrand,
   Presented at 17th International Symposium on Lattice Field Theory 
   (LATTICE 99), Pisa, Italy, 29 Jun - 3 Jul 1999, hep-lat/9907028. } 
\bibitem{la99}{ K.~Langfeld, M.~Engelhardt, H.~Reinhardt and O.~Tennert,
   Presented at 17th International Symposium on Lattice Field Theory 
   (LATTICE 99), Pisa, Italy, 29 Jun - 3 Jul 1999, 
   hep-lat/9908026. } 
\bibitem{creu80}{ M.~Creutz, Phys. Rev. {\bf D21} (1980) 2308. } 



\end{thebibliography}
\end{document}